# MODIFIED CURRENT INDUCED DOMAIN WALL MOTION IN GAMNAS NANOWIRE

N.Vernier[1], J.P. Adam[2], A. Thiaville[2], V. Jeudy[2], A. Lemaitre[3], J.Ferré[2], G. Faini[3]

[1] *Institut d'Electronique Fondamentale, bât. 220, UMR CNRS 8622, Universite Paris-Sud, 91405 Orsay, France*

[2] *Laboratoire de Physique des Solides, bât. 510, UMR CNRS 8502, Universite Paris-Sud, 91405 Orsay, France*

[3] *Laboratoire de Photonique et de Nanostructure, CNRS, route de Nozay, 91640 Marcoussis, France*

**Abstract**

We report on current induced domain wall propagation in a patterned GaMnAs microwire with perpendicular magnetization. An unexpected slowing down of the propagation velocity has been found when the moving domain wall extends over only half of the width of the wire. This slowing down is related to the elongation of a longitudinal wall along the axis of the wire. By using an energy balance argument, the expected theoretical dependence of the velocity change has been calculated and compared with the experimental results. According to this, the energy associated to the longitudinal domain wall should change when a current passes through the wire. These results provide possible evidence of transverse spin diffusion along a longitudinal domain wall.

## Introduction

The propagation of magnetic domain walls with spin-polarized currents has been the focus of intense research over the past decade[1-10]. From a technological perspective, the control of domain walls using electrical currents underlies a number of promising applications in magnetic data storage and logic. From a fundamental perspective, such wall motion is interesting because it involves the interplay between complex spin-dependent transport processes and nonlinear magnetization dynamics.

One key difference between magnetic field and current-driven wall propagation involves the nature of the pressure exerted on the domain wall. For the case in which field-driven magnetization reversal occurs through domain wall propagation, the pressure exerted is always locally perpendicular to the domain wall, irrespective of the orientation of the domain wall. For example, the nucleation of a circular reversed domain in an isotropic film is followed by a circular growth of this domain, where the wall speed is uniform everywhere along the boundary of this domain. In contrast, wall propagation induced by spin-polarized currents depends very much on the direction of the current flow, where the pressure exerted on the domain is largely determined by the scalar product between the current direction and the wall orientation. However, such issues have not been addressed in detail because most experiments to date have been conducted on wire geometries in which the current flow is almost exclusively perpendicular to the domain wall.

Here, we examine the effect of spin-polarized currents on nonuniform domain walls in micron size wires made of the dilute magnetic semiconductor GaMnAs[11-13]. In this material, current-driven wall motion in zero applied fields is known to be possible and the current densities required are typically three orders of magnitude lower than those needed in ferromagnetic transition metals[14-17]. As such, experimental studies on this material are less prone to artefacts arising from sample aging or Oersted fields. We show that current-induced motion velocity in our wires is dependent on how the reversed domain is initialized. By using an analytical model developed and basic energy considerations, it is shown that these feature is a signature of the domain wall interfacial energy. This parameter presents an interesting current dependence discussed at the end.

## Samples, setup and experimental procedure

We have studied current-driven domain wall dynamics in 50 nm thick $Ga_{0.93}Mn_{0.07}As$ layers grown on GaInAs, which exhibit a uniaxial magnetic anisotropy perpendicular to the film plane. The geometry consists of a nucleation pad of size $50 \times 60$ µm$^2$, connected to 6 wires of length 90 µm and of different widths (Figure 1). In the experiments presented here, only the widest wire of width 4 µm has been used. The nucleation of domain walls is initiated in the pad using magnetic pulses. The sample was cooled in an optical open cycle liquid nitrogen cryostat, down to temperatures between 90 K and 130 K. After patterning, the Curie temperature of the sample was found to be 115 K. Other details concerning current-induced domain wall propagation in GaMnAs wires have been presented in previous papers[12,16].

In general, it is important to account for Joule heating in measurements involving current-induced domain wall propagation. The following two-step procedure was employed to obtain the sample temperature during our experiments. First, the resistance of the wire was measured as a function of temperature using a very low applied current (I = 10 µA), which serves as a calibration curve. Next, the resistance of the wire was measured as a function of current for a fixed cryostat temperature $T_0$. By using the initial calibration curve, it is possible to deduce the change in temperature of the wire as a result of Joule heating, $T_0 + \Delta T$, from changes in the resistance of the wire as a function of current $I$. As Figure 2 shows, we observe a parabolic increase in the temperature as a function of current, as expected. This variation has been verified to be reproducible for a number of cryostat temperatures $T_0$. In the following, it is assumed that the temperature attains a steady-state value during the current pulses applied, which allows a good estimate of the sample temperature to be obtained using the calibration method described above. For the results presented here (current densities of $j < 4 \times 10^9$ A m$^{-2}$, pulse durations of $\Delta t \in$ [2 µs, 50 µs]), the temperature underestimation is lower than 0.5 K[18].

The evolution of the magnetization of the sample was tracked by polar Kerr microscopy with a spatial resolution of 2 µm. The sample was illuminated using a red LED. The images were always acquired in the remanent state in zero field. In the temperature range considered in this study, the electrical resistance of the 4µm wide wire was around 16 kΩ. The amplitude of the pulse could be chosen between 0.1 V and 20 V, and its duration varied between 1 µs and 50µs.

**Experimental results**

On figures 3.a and 4.a, one can see a very specific domain wall propagation due to spin polarized current : succesive snapshots of the magnetic state of the 4 µm wide wire are displayed on these figures. The very first picture on the left is a raw image of the wire. The following ones are magnetic pictures obtained by substracting a saturated reference to the actual snapshots. Between each magnetic picture, a current pulse was sent in the wire. It can be seen that the domain walls have moved after each pulse. In the case of picture 4.a, the current used was quite low, so the Joule heating is very low and both cryostat reference and real instantaneous sample temperature were almost equal. The downward domain walls extend over the whole section of the wire, they are the reference for what can be expected for a domain wall extending over the full width of the wire. Let's note that the average velocity of two neighbouring domain walls have been found equal to the one obtained with a single wall and the movement of the first domain walls appear as a good reference[16].

On the upper part of the wire, a new domain appears on the third magnetic image of figure 3.a. This domain has a different morphology, it occupies only half of the wire section, and it is limited by two walls : one perpendicular to the axis of the wire (transverse wall) and one parallel to this axis (longitudinal wall). It can be noted that the size of the reversed domain is one half of the wire width, corresponding to an optimum of the demagnetizing energy. When a current pulse is injected in the wire, the transverse wall moves forward and the longitudinal one gets longer. It must be reminded that a magnetic field would have increased the size of the domain along all directions, which would have resulted in propagation of the longitudinal wall towards the border of the wire until the whole section the wire has got reversed and the longitudinal wall has disappeared. Then, only a full section transverse domain wall is left. We have checked experimentally several times that the magnetic field induced such a behaviour. With current induced motion, the result is completely different, the longitudinal wall does not disppear, it doesn't move, it stays in the middle of the wire and it gets longer as the transverse wall moves. It must be also emphasized that it is a true propagation phenomenon : it cannot be a heating very near the Curie point, resulting in a demagnetized state stabilized by the magnetic field created by the current[19-21].

The second interesting fact is the velocity of the transverse wall : it can be seen on figures 3.a and 4.a that the velocity of the transverse wall decreases when a longitudinal wall elongates. Qualitatively, such a behaviour can be explained by the energy required to elongate the longitudinal domain wall : in addition to the energy necessary to overcome the Gilbert damping for propagating the transverse wall, the current-induced torques must provide some more energy because of the longitudinal wall, so propagation requires more power and becomes slower.

### Quantitative analysis

The slowing down of the transverse wall can be quantitatively calculated as follows. For a domain wall in steady state movement, the velocity can be deduced from an energy balance based on the Landau-Lifschitz-Gilbert equation with the Zhang-Li model for in-plane spin torque[8,22,23]. The power dissipated locally by a torque $\mathbf{\Gamma}$ is given by $\mathbf{\Omega}\cdot\mathbf{\Gamma}$, where $\mathbf{\Omega}$ is the local rotation vector defined by $\partial \mathbf{M}/\partial t = \mathbf{\Omega} \times \mathbf{M}$. The torques applied locally are given by the terms of the Landau-Lifschitz-Gilbert equation in addition to the adiabatic and non-adiabatic contributions due to the spin polarized current, divided by the gyromagnetic ratio $\gamma$. Then, for a wall motion at speed v, the power $\Pi$ dissipated locally is :

$$\Pi = \frac{\partial \mathbf{M}}{\partial t}\cdot \mathbf{B} + \frac{\alpha M \Omega^2}{\gamma} + \frac{\beta u M \Omega^2}{\gamma v} \tag{1}$$

where M is the amplitude of the magnetization, $\alpha$ the Gilbert damping constant and $\beta$ the non adiabatic current induced torque constant. u is the effective spin-drift velocity given by $u = g\mu_B jP/(2q_e M)$, where g is the Landé factor, $\mu_B$ the Bohr magneton, j the current density, P the spin polarisation of the charge carrier and $q_e$ the charge of the electron. Let's note that the power dissipated by the adiabatic term is zero, because $\Omega$ is orthogonal to $\partial \mathbf{M}/\partial t$. By assuming a domain wall profile of $\theta = 2 \arctan[\exp(x-vt)/\Delta]$ and by integrating this power over the whole volume of the transverse domain wall, we obtain the following expression for the power in the flow regime at velocity v :

$$\Pi_T = -2MBvew + \frac{2\alpha M v^2 ew}{\gamma \Delta} - \frac{2\beta u M v e w}{\gamma \Delta} \qquad (2)$$

where e is the film thickness and w the width of the transverse domain wall (figure 3.c). In addition to these terms, in the present case, we have to add the power required to elongate the longitudinal domain wall :

$$\Pi_L = v(\sigma_{DW} e - \sigma_{demag} e) \qquad (3)$$

where $\sigma_{DW}$ is the interfacial energy of the domain wall and $\sigma_{demag}$ the demagnetizing energy. Since wall movement is induced only by applied current in the present experiment, there is no applied magnetic field and the first contribution in $\Pi_T$ is zero. The power provided by the remaining torques should be equal to the power stored by elongating the longitudinal domain wall. This leads to the following solution for the velocity (which can also be directly obtained by realizing that the energy variation due to the change of length of the longitudinal wall has the same form as the energy of the transverse wall under an applied field) :

$$v = \frac{\beta u}{\alpha} - \frac{\gamma \Delta}{2w\alpha M}(\sigma_{DW} - \sigma_{demag}) \qquad (4)$$

By assuming $\sigma_{DW} - \sigma_{demag} > 0$ (ie $\sigma_{DW}/\sigma_{demag} > 1$), the expression obtained agrees with the observed behavior.

These two energies can be determined independently. $\sigma_{DW}$ can be deduced from the spontaneous demagnetized state, which could be observed in the nucleation pad at 104 K (figure 4). However, the observed state is far from the ideal demagnetized structure consisting of perfect stripe domains[24-26], and the relevant stripe periodicity for comparison with the usual formula is not obvious. From figure 4, we have assumed that the periodicity lies between the smallest and the largest width of the observed ribbons, which are 2 µm and 6 µm respectively. By using the values[12] e = 50 nm, $M_S = 15 \times 10^3$ A.m$^{-1}$, we have found the product $e\sigma_{DW}$ to be in the range of $0.85 \times 10^{-12}$ J.m$^{-1}$ to $1.1 \times 10^{-12}$ J.m$^{-1}$. For $\sigma_{demag}$, the usual expression for the demagnetization factor cannot be applied directly because the infinite-film approximation is not suitable to describe the wire. However, the calculation is actually easier because the system is of finite width. The integration of the magnetostatic energy with one domain wall located in the middle of the wire gives the following expression :

$$\sigma_{demag} = \frac{\mu_0 M_s^2}{2\pi} \left\{ 2w\left(2\mathrm{Arctan}\frac{w}{2e} - \mathrm{Arctan}\frac{w}{e}\right) + \frac{w^2}{2e}\ln\frac{w^2+4e^2}{w^2+e^2} + \frac{e}{2}\ln\frac{e^2+w^2}{e^2} - 2e\ln\frac{4e^2+w^2}{4e^2} \right\} \quad (5)$$

Numerically, one obtains $e\sigma_{demag} = 1.0 \times 10^{-12}$ J.m$^{-1}$. This result is compatible with the slowing down of the propagation, as $\sigma_{DW} - \sigma_{demag} > 0$ fits in the range found numerically. At last, from formular (4), in the opposite case for which the length of the longitudinal wall is reduced instead of elongated during the movement, one expects an increase of the velocity. Indeed, on figure 4.b, it has been possible to see such a movement. From this, we could get an approximative value for the velocity of 0.57 m.s$^{-1}$, which is indeed faster than the value 0.49 m.s$^{-1}$ found for full section transverse domain wall, with the same density of the current (figure 4.a).

A careful analysis of the data reveals an interesting fact related to (4): it predicts the difference $v - \beta u/\alpha$ to be constant. However, care must be taken in interpreting the experimental data, since a number of measurements have been acquired in the creep regime of wall motion, while (4) strictly applies only to the flow regime in a perfect system with no defects. In Fig. 6a, we present the experimental wall velocity found in the two cases, i.e., for transverse walls extending across the entire cross-section of the wire and for those that extend across only half the wire. In order to make a meaningful comparison to (4), we deduce the ideal flow velocities v* graphically since these velocities give a better measure of wall motion in perfect samples with no defects. The expected behavior is given by the linear dependence indicated by the solid line in Fig. 6b, whose slope is determined from data obtained at large applied currents[16]. This dependence supposes that the velocity is simply proportional to the pressure exerted on the domain wall. In the case where the transverse wall extends across the full cross section of the wire, this pressure is expected to be proportional to the current density in the absence of applied fields. For the other case where the transverse wall extends across only half of the cross section of the wire, there is an additional contribution due to the elongation of the longitudinal domain wall. From the measured velocity $v_{meas}$, using the graphic procedure presented on figure 6.b, we can get a corrected ideal flow velocity v*.

In Fig. 6c, we show the difference $\Delta v^*$ between the two velocities v* obtained using the procedure described above. Contrary to the behavior described in (4), $\Delta v^*$ is not constant but increases with the current density *j*. How can this behavior be explained? We can discard the Oersted field associated with the current flow as a possible mechanism, since the reversed domain has been observed to occur with equal probability on the left or right part of the wire, despite the current being always applied in the same direction. Furthermore, the magnitude of the Oersted field is expected to attain a maximum of 44 µT at the wire edges, which would only lead to a change in velocity of 0.15 m s$^{-1}$ if this field were applied uniformly across the sample[16], which is not the case.

A more likely explanation of this behavior involves the change in the domain wall energy as the current flows through the wire, which depends on the current density. One mechanism could involve the electrical charging at the wire edges and on the longitudinal domain wall as a result of the anomalous Hall effect, which would lead to an additional electrostatic energy. However, a rough calculation shows that this mechanism is unlikely to be dominant; for this sample, the transverse resistivity[12] $\rho_{xy}$ is found to be lower than $10^{-8}$ $\Omega$.m, which is several orders of magnitude below what is required to explain the experimental result. A more plausible mechanism relates to a change of the domain wall width and energy. Indeed, $\sigma_{DW}$ depends strongly on the wall width and it attains a minimum for the equilibrium width. As it is well known from the Drude model of electrical conductivity in metals, or from the more elaborate Fuchs-Sondheimer calculation of current distribution close to surfaces using the Boltzmann equation, carriers are randomly moving transversely to the electric field, with the mean free path being the characteristic length. On both sides of the longitudinal wall, the majority carriers are of opposite spin polarizations. As a result, the spin transfer torque on this segment of the wall, due to such transverse motion, leads to a variation in the width of this wall. More precisely, in the case of GaMnAs where carriers are polarized opposite to the majority (localized) spins, we anticipate that the width of the longitudinal domain wall increases. As the domain wall width was found to be between 5 and 10 nm in this material[27], and a 40 nm mean free path in GaMnAs may be estimated, this process could be relevant. Although a full calculation of this torque is still lacking, our experiment could provide the first evidence for a large effect of the spin transfer torque due to the transverse motion of the carriers.

## 5. Conclusions

We have reported on current-induced domain wall propagation involving the elongation of a longitudinal domain wall in GaMnAs wires. Such propagation dynamics is peculiar to current-induced motion because the pressure exerted on the domain wall by spin-polarized currents is highly anisotropic. Due to the elongation of the longitudinal domain wall of the reversed domain, the propagation of the transverse wall bounding the domain becomes slower. By using an energy balance argument based on the Landau-Lifshitz-Gilbert equation, it has been possible to determine the expected behavior for the propagation velocity in this scenario. By examining the demagnetized state, it was possible to deduce the ratio between the demagnetized energy and domain wall interfacial energy, which agrees with our expectations. A careful analysis of the change in velocity between the cases with and without longitudinal wall has lead to possible evidence of wall width changes due to transverse spin diffusion. Such processes are important because they can appear in any material with perpendicular magnetic anisotropy. For example, in FePd nanowires, the demagnetized ribbon periodicity is very small[28], and they could occur in wire as narrow as 100 nm. GaMnAs is a very good model system as it is possible to study this kind of phenomenon using the efficiency of Kerr microscopy, while the smaller domain sizes in FePd would require more sophisticated tools.

The authors wish to thank Joo-Von Kim for his comments. This work was supported by the MANGAS project (2010-BLANC-0424) from the French National Research Agency (ANR).

# References (first version)

Figure captions

Figure 1 : optical view of the GaMnAs wires and nucleation pad, with the contact pads and the bonding wires.

Figure 2 : temperature increase as a function of the current density in steady state, the continuous line is a fit made with the parabolic law $T = T_0 + 6.27 \times I^2$ (I being in mA) as expected for a heating due to the Joule effect. Inset curve : temperature dependence of the nanowire resistance, the dotted line is the extrapolation function used to determine T as a function of R.

Figure 3 : (a) set of 12 pictures of the nanowire. The size of each picture is $18 \times 98 \mu m^2$. The very first one on the left is the optical reference picture of the nanowire in a saturated magnetic state. The 11 other pictures show the magnetic states obtained through Kerr microscopy of the studied nanowire. Between each picture, a current pulse of amplitude $2.1 \times 10^9$ A.m$^{-2}$ was sent. The duration of the pulse was 2.4 µs. The cryostat temperature was 101.1 K and the expected steady sate temperature of the wire was 104 K. (b) is a sketch of the magnetic pictures observed and the walls limiting the domains. The dash lines represent the limits of the nanowires, which cannot be seen on the magnetic pictures. $\otimes$ means a magnetization pointing down and $\odot$ a magnetization pointing up. The dot line holds for the domain walls refered as "half section transverse domain wall" the full lines hold for the domain walls refered as "full section transverse domain walls", and the mixed dash-dot line holds for the "longitudinal domain wall".

Figure 4 : .(a) and (b) are a set of pictures of the nanowire. The very first picture on the left of each set (a) and (b) is the raw optical picture used as a reference. The other ones are magnetic pictures obtained through Kerr microscopy of the studied nanowire. The last picture of set (a) has been reported on set (b) and is the first magnetic picture from the left. Between each picture, a current pulse of amplitude $0.73 \times 10^9$ A.m$^{-2}$ was sent. For set (a), the duration of the pulse was 15 µs, for set (b) , the duration was 50 µs. The cryostat temperature was 103.6 K and the expected steady sate temperature of the wire was 104 K. (c) .

Figure 5 : Kerr picture of the demagnetized state at 104 K. It was acquired on the nucleation pad, the size is $40 \times 40 \mu m^2$.

Figure 6 : (a) velocity for the two types of domain wall as a function of the density of current, (b) graphical procedure to get the ideal velocity v* expected for a perfect sample without defect (c) difference of v* for the types of domain walls as a function of the density of current.

Figure 1

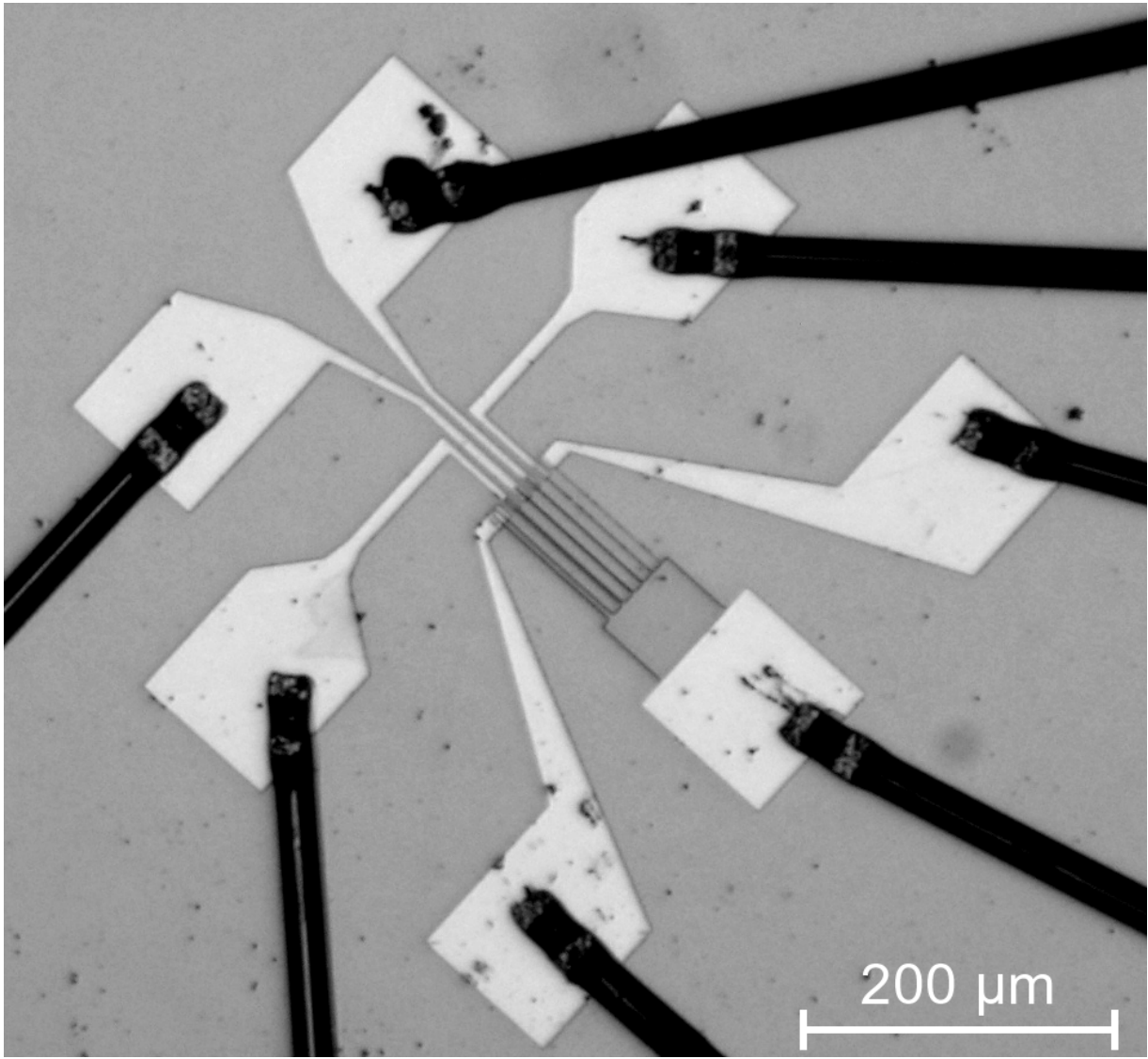

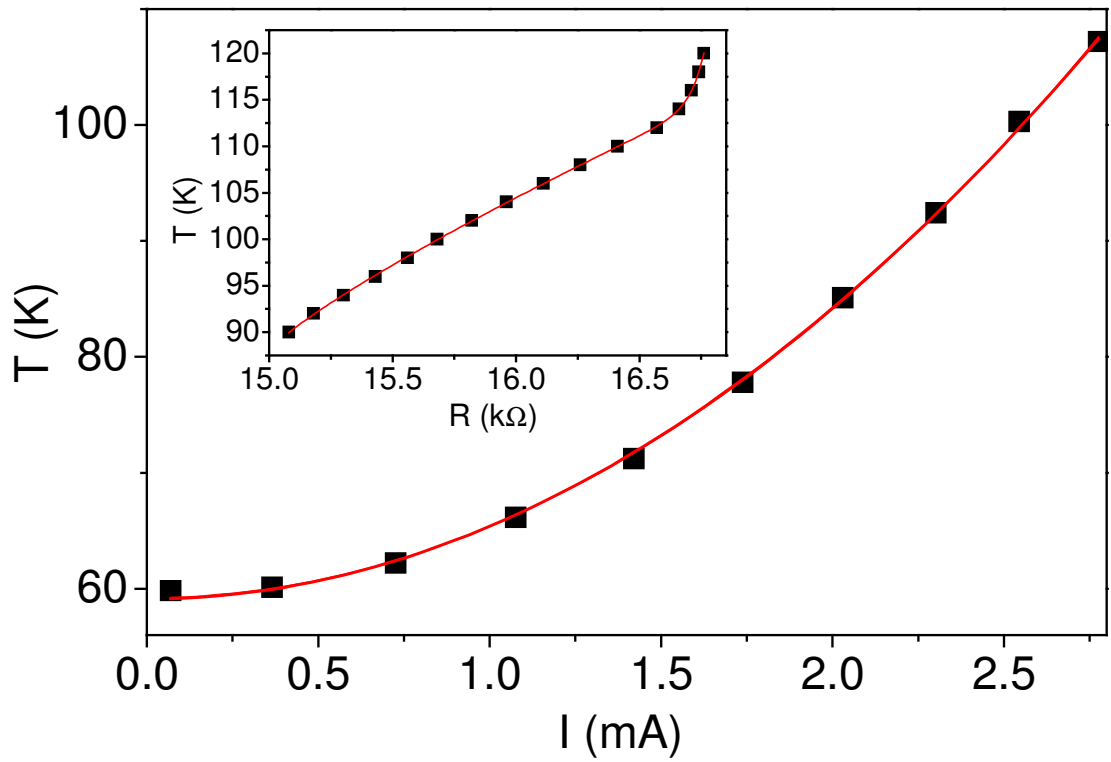

Figure 3

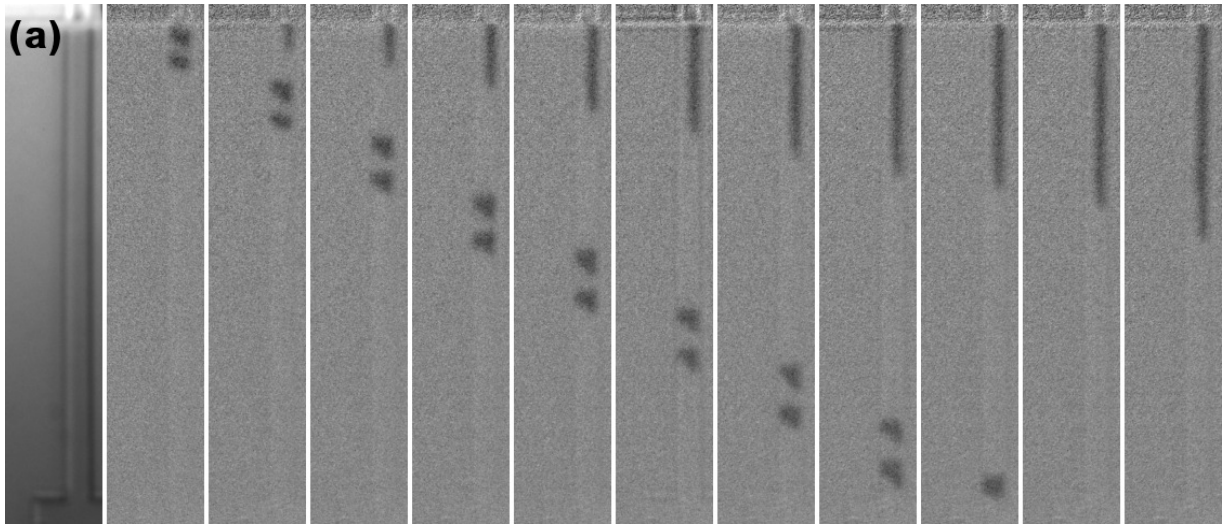

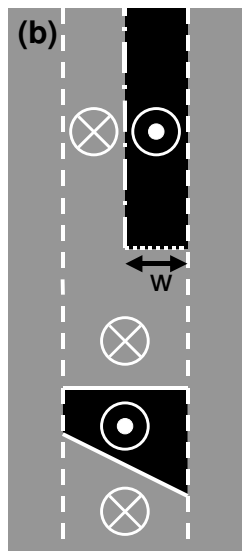

Figure 4

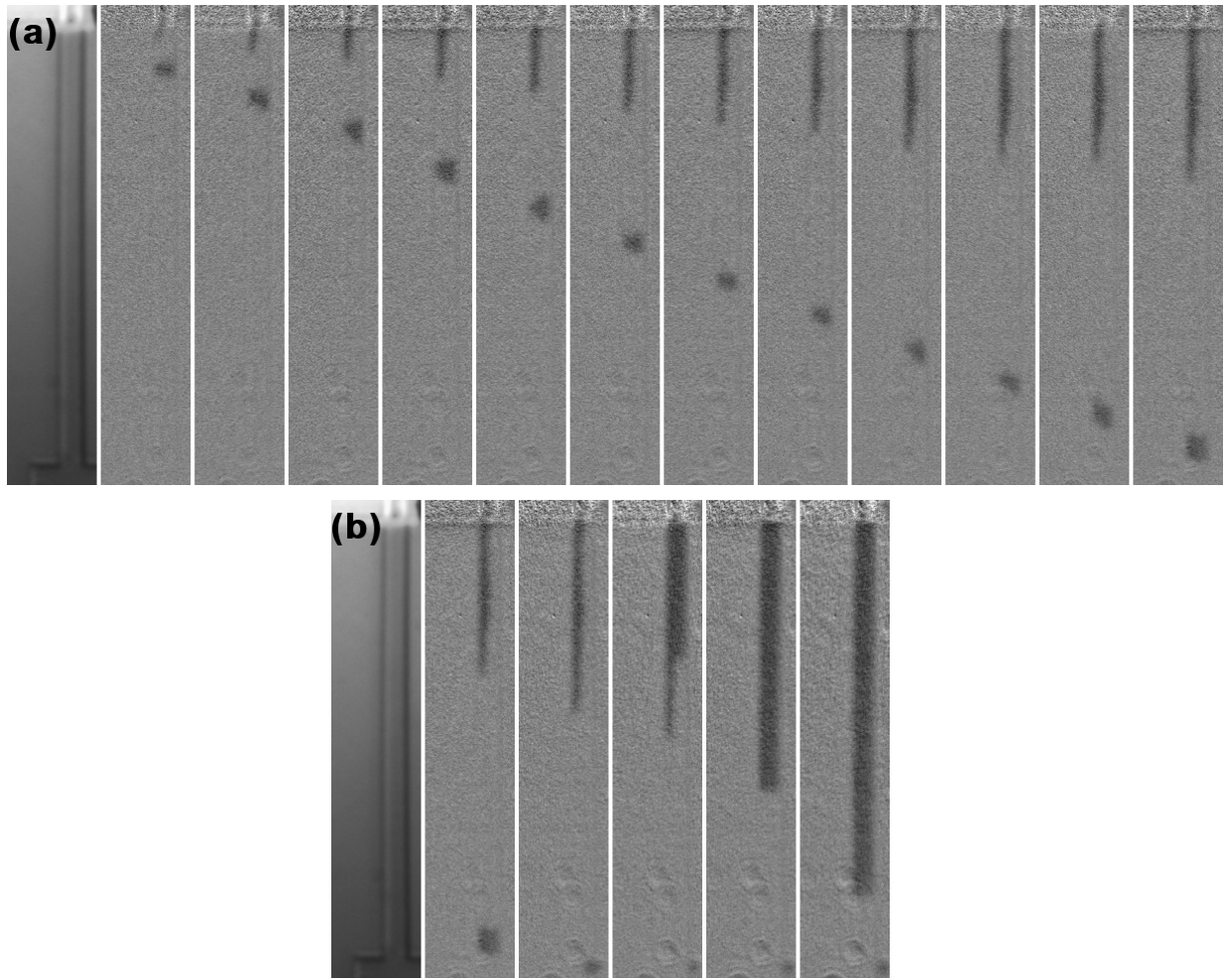

Figure 5

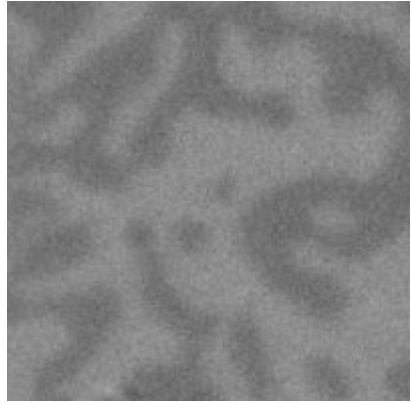

Figure 6

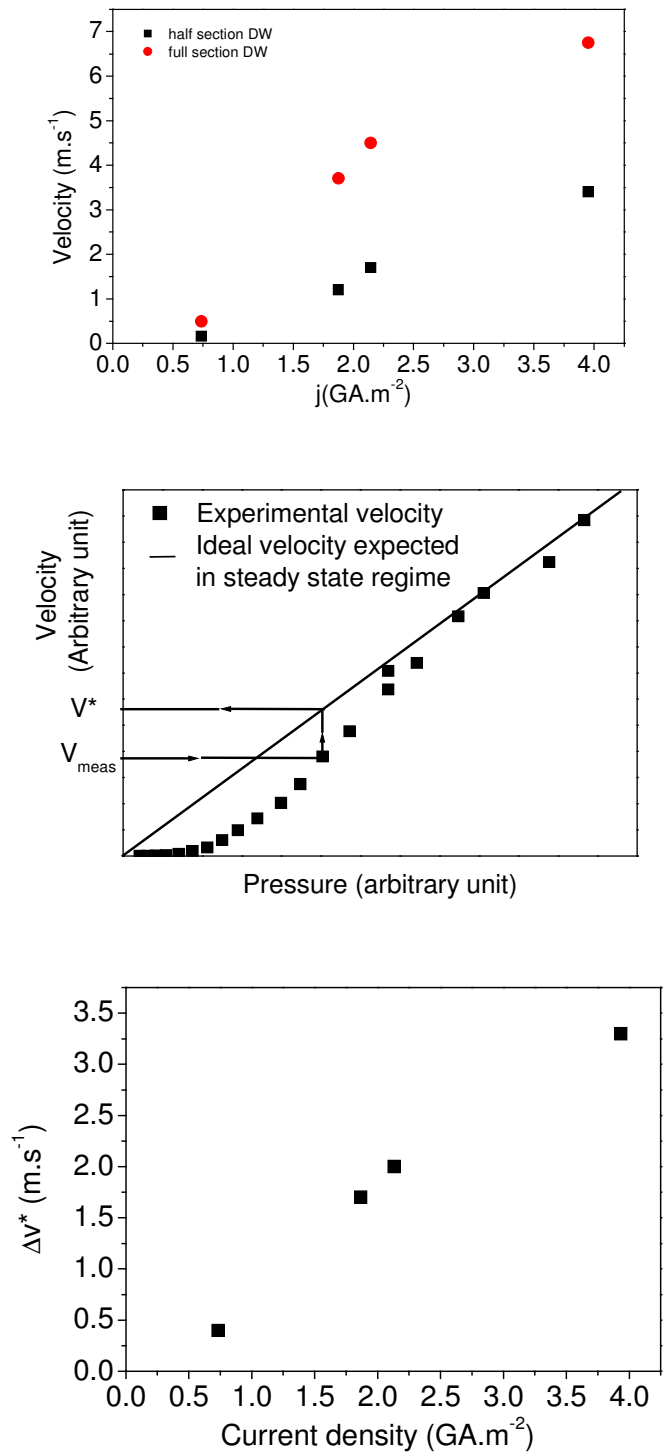